\documentclass[%
 reprint,
 amsmath,amssymb,
 aps,
pra,
superscriptaddress,
]{revtex4-1}

\usepackage[english]{babel}
\usepackage{amsmath,amssymb}
\usepackage{setspace}
\usepackage{hyperref} 
\usepackage[utf8x]{inputenc}
\usepackage{caption}
\usepackage{leftidx}
\usepackage{graphicx}
\usepackage{epstopdf}
\usepackage{dcolumn}
\usepackage{bm}
\usepackage{float}

\begin{document}

\preprint{APS/123-QED}

\title{Quantum Memory and Optical Transistor Based on Electromagnetically Induced Transparency in Optical Cavities}

\author{Rommel R. Oliveira}
\affiliation{Departamento de F\'{i}sica, Universidade Federal de S\~{a}o Carlos, P.O. Box 676, 13565-905, S\~{a}o Carlos, S\~{a}o Paulo, Brazil\\}
\author{Halyne S. Borges}
\affiliation{Departamento de F\'{i}sica, Universidade Federal de S\~{a}o Carlos, P.O. Box 676, 13565-905, S\~{a}o Carlos, S\~{a}o Paulo, Brazil\\}
\email{halyneborges@gmail.com}
\author{James A. Souza}
\affiliation{Departamento de F\'{i}sica, Universidade Federal de S\~{a}o Carlos, P.O. Box 676, 13565-905, S\~{a}o Carlos, S\~{a}o Paulo, Brazil\\}
\affiliation{Departamento de F\'{i}sica, Qu\'{i}mica e Matem\'{a}tica, Universidade Federal de S\~{a}o Carlos, 18052-780, Sorocaba, S\~{a}o Paulo, Brazil\\}
\author{Celso J. Villas-Boas}%
\affiliation{Departamento de F\'{i}sica, Universidade Federal de S\~{a}o Carlos, P.O. Box 676, 13565-905, S\~{a}o Carlos, S\~{a}o Paulo, Brazil\\}
%
%
%

\date{\today}

\begin{abstract}
We theoretically studied the implementation of a quantum memory and an optical transistor in a system composed by a single atom trapped inside a high finesse cavity. In order to store and map the quantum state of an input pulse onto internal states of the single atom (quantum memory) we employ the electromagnetically induced transparency (EIT) phenomenon (which can work out as an optical transistor) where the information can be transferred to the dark state of the atom modelled by a three-level system in a $\Lambda$-type configuration. In our model we consider a suitable temporal shape for the control field that ensures the adiabaticity of the storage process and retrieval of the probe pulse. The dynamic of the field inside the cavity was obtained by master equation approach, while the outside field was calculated by input-output formalism. We have analysed two different setups: i) two-sided and ii) single-sided cavities. While the first setup is the most appropriate and commonly used to observe cavity-EIT in the transmission spectrum, thus the best configuration for the optical transistor, the maximum quantum memory efficiency can not reach reasonable values, being limited to $50\%$ for symmetric cavities. On the other hand, with single-sided cavity the quantum memory efficiency increases considerably and can reach values close to $100\%$ in the strong atom-field coupling regime. However this specific setup is not favourable to observe the cavity-EIT effect in the transmission spectrum and then it is not appropriate to control the transmission of light pulses. 

\end{abstract}

\maketitle


\section{Introduction}
\label{sec:1}

The implementation of quantum information processing requires the ability to perform different tasks with high efficiency and control, for instance the initialization and detection of quantum states, storage of quantum information (quantum memory), quantum logic gates, control the transmission of information (transistor), etc. The optical transistor \cite{Parkins10, Lukin13} can be defined as a device which is able to control the transmission of light, allowing the light to be either reflected or transmitted through the application of a second light field (control field). In this way, the transmission of quantum information mapped onto a light pulse can be controlled via another light field. Recently this control was achieved using a trapped atom inside an optical cavity \cite{Villas_Boas10}, based on the electromagnetically induced transparency (EIT) phenomenon, opening a great avenue for many applications in quantum information area. In this experiment, even a single atom was able to control the transmission of light of a probe field, although the contrast between the transmission of the probe field when the control field is on and when it is off was of only $20\%$ \cite{Villas_Boas10}. For an average of $15$ atoms this contrast reached much higher values, over $90\%$. Another essential device for quantum information processing is the quantum memory \cite{Sanders09}, which is a system capable of storing quantum states to perform a certain task. They can be applied not only in quantum computation but also in quantum repeaters, metrology, detection and emission of single photons, quantum networks and as a system to study fundamentals of quantum mechanics \cite{Simon10, Brussieres13}. In general, an optical quantum memory transfers the properties of a photonic qubit input and maps them onto a medium, being retrieved after a storage time with a significant efficiency and fidelity. Among the several physical systems used for its implementation are atomic ensembles in solid state \cite{Philipp13}, systems with a single atom in optical cavities \cite{Rempe11}, atomic gases \cite{Riedl12}, semiconductor quantum dots \cite{Young07} and ensemble of nuclear spins in quantum dots \cite{Fleischhauer09}. Recently a single-atom quantum memory was accomplished with a $^{87}\mathrm{Rb}$ atom, where a photon state given by a superposition of the right and left polarization was stored into a superposition of the atomic states, with efficiency of $9.3\%$ for storage time of $2\mu s$ and an average fidelity of $93\%$ for a storage time of $180\mu s$ \cite{Rempe11}. In the context of applications of quantum memory such as quantum repeaters which needs an efficiency of over $90\%$, and in linear optical quantum computation, which needs efficiencies over $99\%$ \cite{Brussieres13}, it is extremely relevant to optimize the memory efficiency. 

In this work we investigate theoretically the implementation of an optical quantum memory and an optical transistor in a system composed by a single atom, modelled as a three-level system in $\Lambda$-configuration, trapped in a high finesse optical cavity. The state of the probe pulse is coherently mapped onto the atomic levels, considering the cavity in two different setups, whose distinction is given by the difference between the reflectivity of each one of the mirrors. Here we have investigated several ways to optimize the quantum memory efficiency considering different parameters values, which in turn can be properly controlled experimentally. Based on the nonlinear optical phenomenon EIT, the state of a photonic qubit is transferred to a superposition of the two atomic ground states. This storage process is accomplished turning the control field off adiabatically, ensuring that the information of the input qubit remains stored in an eigenstate of the system that does not contain any contribution of the excited state, i.e., in the dark state. 

Our results show that for a two-sided cavity, which is the appropriate experimental setup to observe cavity-EIT effect with single atoms \cite{Villas_Boas10}, the value of memory efficiency has an upper bound about $8.5\%$ for the asymmetric setup used in \cite{Villas_Boas10}, not being suitable for quantum memory application. This setup is neither convenient to implement an optical transistor since most of the light is immediately reflected, independently of the control field, as we discuss below. The symmetric two-sided cavity allows for $100\%$ transmission of light, being the most suitable setup to implement the optical transistor since one is able to obtain $100\%$ ($0\%$) transmission when the control field is on (off) in the limit of strong atom-field coupling. However, here we show that this septup provides a quantum memory efficiency limited to $50\%$.  Meanwhile, the one-sided cavity setup has its maximum value of the memory efficiency significantly increased in relation to the last configurations, close to $100\%$, however it is not suitable to observe the cavity-EIT. In this specific configuration, the reflected and transmitted fields become indistinguishable under transmission measurements. So, the one sided-cavity setup is not useful for the implementation of the optical transistor. 


The manuscript is divided as follows. In Sec. \ref{sec:2} we present the theoretical model. In Sec. \ref{sec:3} we investigate the optimization of the quantum memory efficiency and discuss the implementation of the optical transistor  for the two different setups. Finally we present our conclusions in Sec. \ref{sec:4}.

\section{Theory and Model}
\label{sec:2}
In order to investigate the dynamics of the atom-field system inside the cavity we calculate the density matrix via master equation formalism. The Hamiltonian that describes the atom-cavity system under the influence of probe and control fields, in the rotating wave approximation and without the time dependency, is given in the interaction picture by: 
\begin{equation}
\begin{split}
H_{I}=& {\Delta _{1}\sigma _{33}+(\Delta _{1}-\Delta _{2})\sigma
	_{22}+\Delta \sigma _{11}-\Delta a^{\dagger }a+} \\
& {(\varepsilon a+ ga\sigma_{31}+\Omega _{C}\sigma _{32}+ h.c.)}\text{.}
\end{split}
\label{eq:1}
\end{equation} 
Here, $\Delta _{1}=\omega _{3}-\omega $ is the detuning between the $%
\left\vert 3\right\rangle \leftrightarrow \left\vert 1\right\rangle $ atomic
transition and the cavity field frequencies, $\omega _{3}$ and $\omega $,
respectively. $\Delta _{2}=(\omega _{3}-\omega _{2})-\omega _{C}$ is the
detuning between the $\left\vert 3\right\rangle \leftrightarrow \left\vert
2\right\rangle $ atomic transition ($\omega _{3}-\omega _{2}$) and the
control field ($\omega _{C}$) frequencies. $\Delta =\omega _{P}-\omega $ is
the detuning between cavity mode ($\omega$) and probe field ($\omega _{P}$)
frequencies. $\sigma _{kl}=\left\vert k\right\rangle \left\langle
l\right\vert $ ($k,l=1,2,3$) are the atomic operators: for $k=l$ we have the
population operators and for $k\neq l$ the transition ones. The cavity couples the $|3\rangle \leftrightarrow |1\rangle $
transition with a coupling constant $g$ and the control laser, with Rabi frequency 
$\Omega _{C}$, couples the $|3\rangle \leftrightarrow |2\rangle $
transition. $h.c.$ stands for Hermitian conjugate. The operators $a$ and $a^{\dagger}$ are associated to the internal cavity mode. Finally, $\varepsilon$ is the strength of the probe field on the cavity mode. Concerning the atom-cavity coupling is important to point out that we do not take into account oscillations of the atom in the cavity, i.e., we do not consider any deviations in the
value of the coupling constant $g$, considering in all our results an effective coupling.

\begin{figure}[b]
\centering
\includegraphics[width=1.0\linewidth]{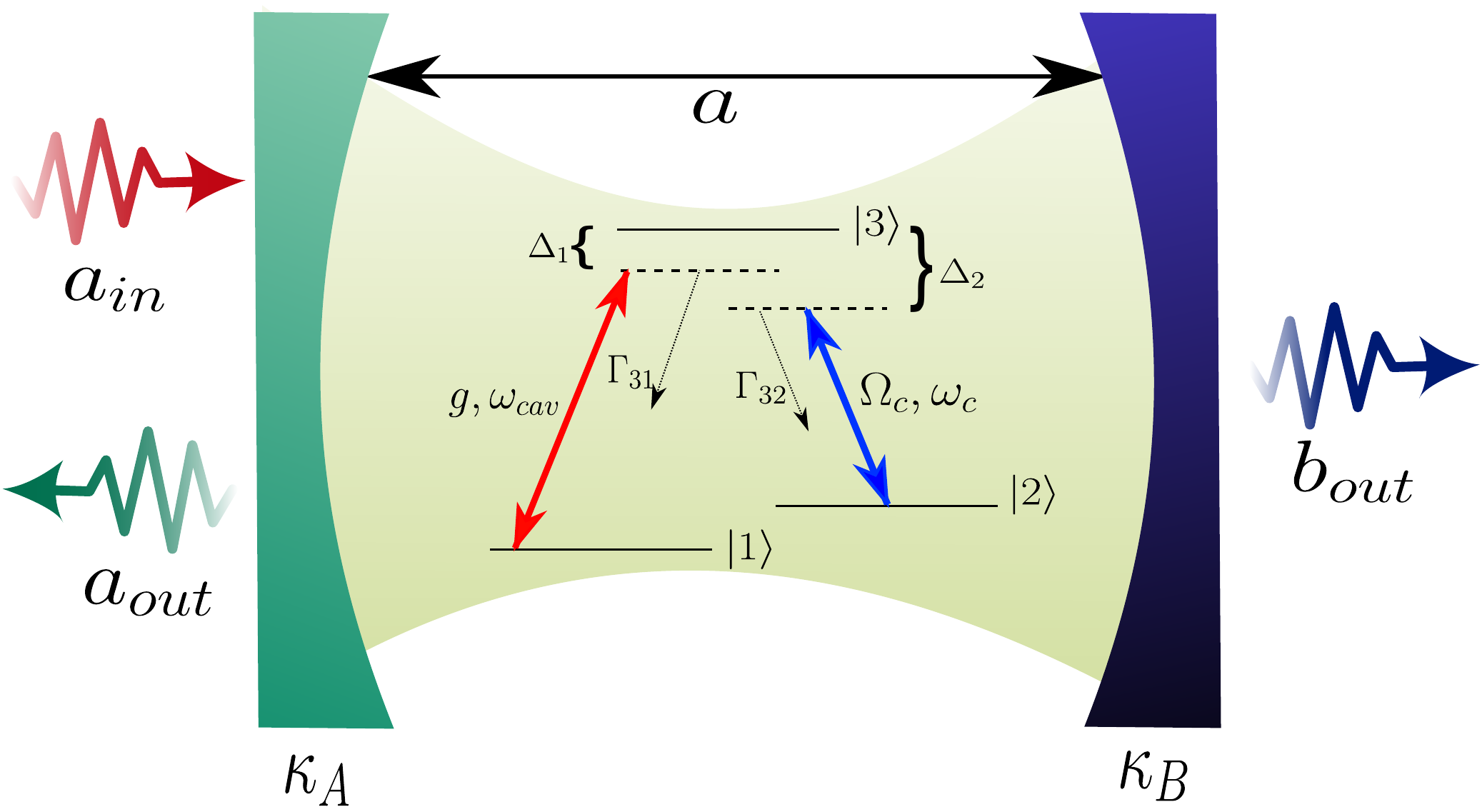}  
\caption{Energy level diagram for a single atom in the $\Lambda $ configuration trapped inside a high finesse cavity. 
The incoming field on the cavity is represented by $a_{in}$. The cavity couples the $|3\rangle \leftrightarrow |1\rangle $ transition with a coupling constant $g$. Meanwhile, a control laser of Rabi frequency $\Omega _{C}$
couples the $|3\rangle \leftrightarrow |2\rangle $ transition. $\Gamma_{31}$ and $\Gamma_{32}$ represent the polarization decay rates from the excited state $|3\rangle$.}
\label{fig:1}
\end{figure}

Fig.\ref{fig:1} shows the schematic representation of the atom-cavity system with couplings and atomic decay rates. The constants $\kappa_A$ and $\kappa_B$ represent the cavity field decay rates associated to each one of the mirrors.

Since we are interested in the implementation of the optical transistor and the quantum memory based on EIT phenomenon, we assume from now on that the control laser is resonant with the atomic transition $|3\rangle \leftrightarrow |2\rangle$, the probe field is tuned to resonance with
cavity frequency, which in turn couples resonantly the atomic transition $|3\rangle \leftrightarrow |1\rangle $. Therefore, all the detunings of the Hamiltonian (\ref{eq:1}) are null. 

The principle behind quantum memories based on EIT is to use the transparency window generated by this phenomenon and, consequently, the reduction of the group velocity of the light pulse into the medium. For the atom-cavity system, first one must prepare it in the ground atomic state $|1\rangle_a$ and the cavity in the Fock state $|0\rangle_c$ (the index $a$ and $c$ refer to the atom and cavity, respectively).  If instead of a continuous probe field, a pulse with one photon is sent, with frequency spectrum within the transparency window, in the limit where $\Omega_C\gg\Omega_P$ (being $\Omega_P$ the Rabi frequency of the probe field), the dark state of the system is given by the tensor product $|1\rangle_a\otimes |1\rangle_c$, since one photon was inserted into the cavity. As the control laser is on, under these conditions the probe pulse is not absorbed by the atom. However, if the control field is adiabatically turned off, keeping the system in an eigenstate of the Hamiltonian (its dark state), the pulse is now absorbed and the dark state of the compose system is now $|2\rangle_a\otimes |0\rangle_c$. Similarly, the input pulse can be a coherent superposition of 0 and 1 photon and, then when the control field is adiabatically turned off, it will be transferred and stored in a superposition of the ground atomic levels. For coherent probe pulses the state of the light is a coherent superposition of 0, 1, 2... photons, which can be well approximated (for quantum memory purposes) to a superposition of 0 and 1 photon for a very weak pulse, as we consider in this work.
 
In order to use this model to describe a quantum memory made up of this system, as carried out by H. P. Specht \textit{et al}. \cite{Rempe11},
we consider a weak coherent probe pulse incoming to the cavity with the following Gaussian temporal shape:
\begin{equation}
\phi_{in} (t)=E_{m}e^{-\frac{1}{2}\frac{(t-t_{0})^{2}}{\alpha ^{2}}}\text{,%
}  \label{eq:2}
\end{equation}%
where its full width at half maximum ($FWHM$) is given by $FWHM=2\alpha 
\sqrt{2\ln (2)}$. $t_{0}$ is the time the probe pulse (its maximum) enters the cavity mode. This probe field is what we are interested in
storing in the atom. In order to do the storage, we also must turn the control field off, making the atom absorb the probe pulse. In the following we have to turn the control field back on, in order to restore the probe field stored in the atom. Turning off and on the control field abruptly one can excite the atomic level $\left\vert 3\right\rangle $, which can decay spontaneously, losing energy and information. To avoid this source of error we must turn off and on the control field adiabatically, which ensures that the system will be kept all the time in its dark state \cite{Rempe99}. In order to ensure the adiabaticity in the storage process, we consider the time dependency to the control field given by: 
\begin{equation}
\begin{split}
\Omega_{C}(t)=& {\frac{\Omega _{C}}{2}\{[1-tanh(\zeta _{1}(t-t_{1}))]+}{[1+tanh(\zeta _{2}(t-t_{2}))]\}\text{,}}
\end{split}
\label{eq:3}
\end{equation}%
where $\zeta _{1}$ ($\zeta _{2}$) is the rate at which we turn the control field off (on),
at time $t_{1}$ ($t_{2}$) and $\Omega_{C}$ is the maximum Rabi frequency of the control field. With this simple form for the control field we are able to reach memory efficiencies close to $100\%$ for the one sided cavity and Gaussian coherent probe pulses. Thus we are not using the protocols by J. Dilley \textit{et al.} \cite{Dilley12} or M. Fleischhauer \textit{et al.} \cite{Lukin00} which maximize the memory efficiency only for single photon pulses (i.e., not being valid for weak coherent probe pulse as we consider here) and at the expense of the derivation of specific forms of the control field for each shape and width of the probe pulse.

In cavity quantum electrodynamics is well established that an optical cavity driven by an electromagnetic field can be described by two formalisms: via master equation, where the behaviour of the field inside the cavity is completely described independently of the cavity design, and by input-output formalism, that describes explicitly the input and output fields that are emitted from or reflected by the cavity, through Heisenberg-picture operators \cite{Gardiner85}. In the case of Fabry-Perot cavities, the physical system can be constituted either by a perfectly reflective mirror, while the other one is partially reflective, in such a way that the field can only enter and exit from cavity by one side (one-sided cavity), or by mirrors which exhibit non null reflection and transmission coefficients so that, it is possible to consider that the input field is sent through one side of the cavity while the output field can exit by both sides (two-sided cavity). As a particular case of two-sided cavity we have the symmetric one, which consists of a cavity where both mirrors have the same transmission coefficients.

In order to calculate the dynamics of the internal field of the atom-cavity system we use the master equation    

\begin{equation}
\begin{split}
\frac{d\rho }{dt}=& -i[H_{I},\rho ]+\kappa (2a\rho a^{\dagger }-a^{\dagger
}a\rho -\rho a^{\dagger }a) \\
& +\sum_{i=1,2}\Gamma _{3i}(2\sigma _{i3}\rho \sigma _{3i}-\sigma
_{3i}\sigma _{i3}\rho -\rho \sigma _{3i}\sigma _{i3}),
\end{split}
\label{eq:4}
\end{equation}%
being $\kappa=\kappa_A+\kappa_B$ the total decay rate of the cavity field, $\Gamma _{32}$ and $\Gamma
_{31}$ the polarization decay rates of the excited level $3$ to levels $2$
and $1$, respectively. 

As we are interested in optimizing the quantum memory efficiency of this system, it is important to emphasize that in order 
to evaluate the memory efficiency is necessary to know the input and output fields, and their corresponding relations with the internal cavity mode (described by the operators $a(t)$ and $a^{\dagger}(t)$) calculated through of the master equation (\ref{eq:4}). 

In this context, the input-output theory provides an important relation between cavity mode and the external modes \cite{Gardiner84}: 
\begin{equation}
a_{out}(t)=\sqrt{2\kappa_A}a(t)-\phi_{in}(t),
\label{eq:5}
\end{equation}
where the operators $\phi_{in}$ and $a_{out}$ describes the incoming and outgoing fields, respectively, for a single sided cavity whose field decays at a rate $\kappa_A$. 

The connection between the input-output formalism and the master equation approach is performed fixing $\varepsilon = -i\sqrt{\kappa_A} \phi_{in}$. 

The extension to a two-sided cavity is straightforward. If we consider the input field $\phi_{in}$ is sent only through one side of the cavity, the output mode $a_{out}$, now related to the light field which leaves the cavity by its left side ($L$), obeys the equation (\ref{eq:5}). Meanwhile, the outgoing field described by the operator $b_{out}$, related to the photons which leaves the right side of the cavity ($R$), is given by the relation: $b_{out}(t)=\sqrt{2\kappa_B}a(t)$, since in our model we consider that the input pulse impinges on the left side of the cavity, i.e. $b_{in}=0$. The main difference between the outgoing fields of each one of the sides of the cavity is an interference process occurring between the input field and the one reflected by the cavity, which does not occur for the transmitted field on the right side.

Here we have investigated the parameter configuration of the system in which the quantum memory efficiency and the optical transistor are optimized considering two different setups. In our model each one of these setups are distinguished by the relations between the decay rates of the cavity field associated to each one of the mirrors. The two different configurations are: two-sided (asymmetric or symmetric) cavity when the cavity decay rates $\kappa_A$ and $\kappa_B$ are non null and we labelled as configuration $\mathrm{I}$; the one-sided cavity, when $\kappa_A$ is non null and $\kappa_B = 0$, is labelled as configuration $\mathrm{II}$. It is important to highlight that the two-sided cavity is the common setup used to perform cavity EIT experiments \cite{Villas_Boas10}. 

\section{Optimizing the quantum memory efficiency and the optical transistor}
\label{sec:3}

In order to analyse the feasibility of implementing a quantum memory and/or an optical transistor in the atom-cavity system based on the EIT phenomenon, we investigated in detail which parameter configuration the memory efficiency/optical transistor are optimized considering the two different setups described previously. Considering our model in which the fields inside and outside the cavity were obtained through of master equation approach (\ref{eq:4}) and input-output formalism respectively, we are able to calculate the memory efficiency and transmission in order to identify the parameter configuration experimentally accessible and investigate how they affect the maximum efficiency value and the performance of the optical transistor. From ours simulations we determined to which parameter configuration associated to the probe pulse $\varepsilon(t)$, whose temporal dependency is given by equation (\ref{eq:2}), and control field $\Omega_C(t)$, expressed by equation (\ref{eq:3}), the value of memory efficiency is optimized.

\textit{Quantum Memory:} As well as in the reference \cite{Rempe11}, here the memory efficiency is defined as the total mean number of retrieved photons from the cavity mode (after turning on the control field) in the presence of the atom, normalized by the input pulse:
\begin{equation}
\eta=\langle n_{out}\rangle_{stored}/\int dt |\phi_{in}(t)|^2,
\label{eq:efficiency1}
\end{equation}
where $\langle n_{out}\rangle_{stored}$ is the mean number of photons associated to the fraction of retrieved photons. For the one-sided cavity the efficiency is provided by calculating the mean number of photons which leave the cavity by its left side, i.e., $\langle n_{out}\rangle_{stored} = \int dt \langle a_{out}^{\dagger}a_{out}\rangle_{stored}$. On the other hand, for the two-sided cavity setup we compute the quantum efficiency taking into account the retrieved photons which leave the cavity by both sides, i.e., in this case $\langle n_{out}\rangle_{stored} = \int dt \langle a_{out}^{\dagger}a_{out}\rangle_{stored} + \int dt \langle b_{out}^{\dagger}b_{out}\rangle_{stored}$. 

\begin{figure}[h]
\centering
\includegraphics[width=1.0\linewidth]{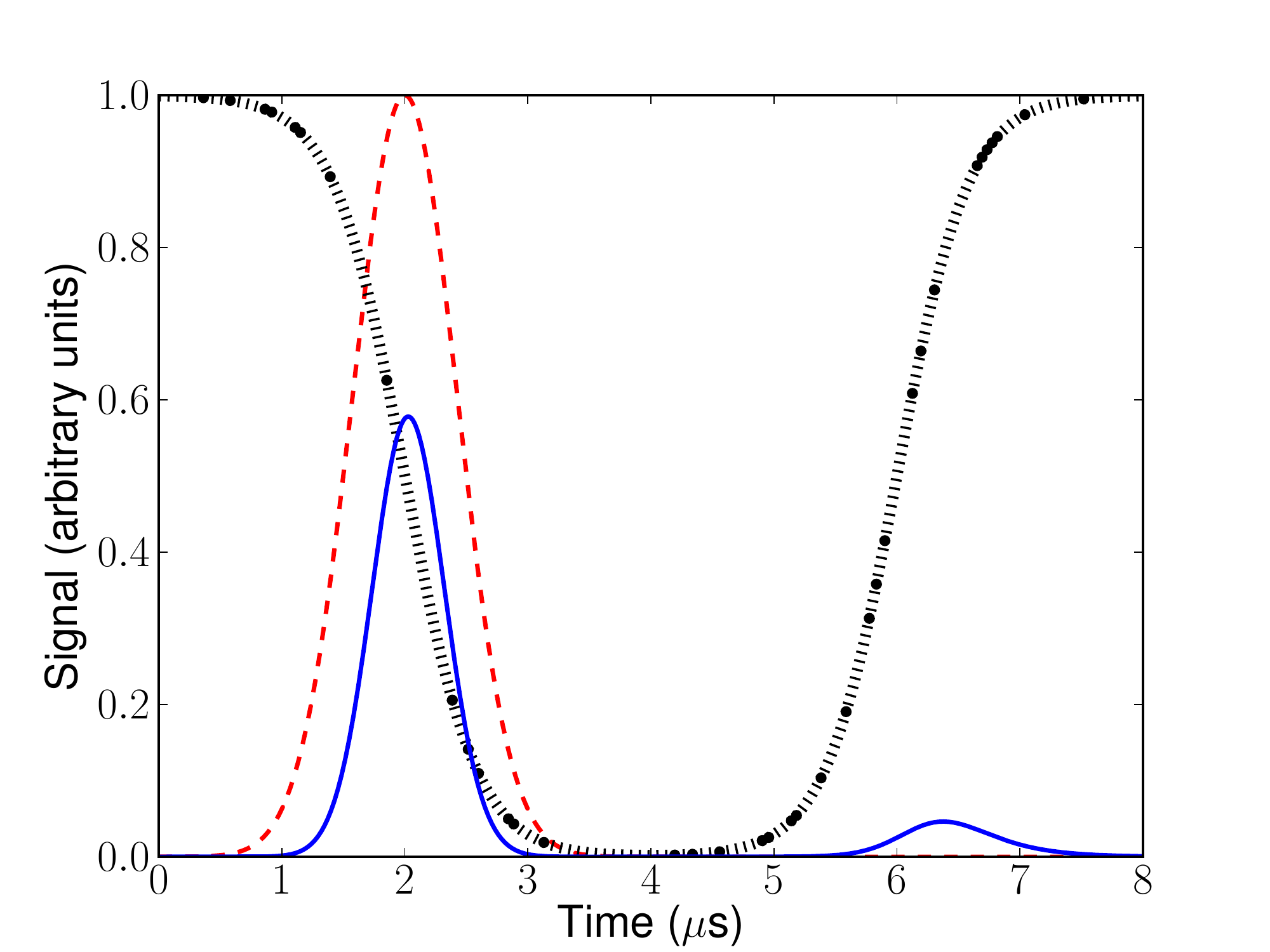}
\caption{Single atom quantum memory scheme considering the configuration $\mathrm{I}$, asymmetric cavity ($\kappa_B=0.1\kappa_A$). The black dotted curve represents the control field $\Omega_C$, the red dashed curve is the input pulse $\varepsilon$ and the solid blue curve is the mean number of photons in the cavity. The parameters used here were: $\kappa/2\pi=2.5$MHz, $\Gamma_{31}=\Gamma_{32}=0.6\kappa$, $g = 1.0\kappa$, $\Omega_C = 2g/3$, $E_M = \sqrt{10^{-4}}\kappa$, $\zeta_1 = \zeta_2 = \zeta = 1.75$MHz, and $FWHM = 1.0\mu s$. We obtained an efficiency of $9.35\%$ }
\label{fig:2}
\end{figure}

Using the simple model described in the previous section we were able to reproduce the results obtained in the experimental work \cite{Rempe11} as is shown in Fig. \ref{fig:2}. The setup used to perform the experiment was the configuration detonated by us as $\mathrm{I}$ (two-sided cavity), but, as $\kappa_B=0.1\kappa_A$ this setup can be approximated by an one-sided cavity. In Fig. \ref{fig:2}, we plot the fields involved in the quantum memory process. The Gaussian pulse is represented by the red dashed curve with its maximum at $t_0=2\mu\mathrm{s}$, the black dotted curve is the temporal form of the control laser turned off and on at $t_1=2\mu\mathrm{s}$ and $t_2=6\mu\mathrm{s}$, respectively. The mean number of photons inside the cavity is given by the solid blue curve, where the second peak whose maximum is around $6.5\mu\mathrm{s}$ corresponds to the part of the retrieved photon pulse from the memory, while the first peak is associated to incident light that is directly transmitted by the left mirror.

It is relevant to mention that in this result we assumed an effective atom-field coupling $g$ as approximately half of the maximum ideal coupling, as it was done in \cite{Villas_Boas10} in order to theoretically fit the experimental data for the cavity EIT with single atoms. This has been done since the atom is not perfectly cooled and then, due to its motion, it is not always in the maximum coupling position. With an effective coupling $g\simeq g_{\max }/2$ we could obtain a theoretical efficiency of $9.35\%$ for restoring the stored pulse, perfectly close the value $9.3\%$ obtained experimentally \cite{Rempe11}. 

For not so large atom-field coupling $g$, one important condition to observe cavity EIT phenomenon is the probe laser with low intensity. In our simulations we consider the probe laser amplitude $E_M$ small enough such that the probability of more than one photon in the cavity is approximately null. From our analysis we obtained, for the amplitude value $E_M=\sqrt{10^{-4}}\kappa$, the efficiency assume its greatest value. Another parameter that influences the maximum efficiency is the time the pulse enters the cavity ($t_0$) relatively to the time we choose to turn the control field off ($t_1$). The fulfilled condition $t_1-t_0=0$, ensures also a best memory efficiency \cite{Lukin00}. In relation to the others parameters associated to the control field, the storage process is optimized when the rates that determine how fast the control field is turned off and on assume an equal value, such that, $\zeta_1=\zeta_2=1.75\mathrm{MHz}$. It is also extremely relevant to investigate the effect of the control laser amplitude $\Omega_C$ and the atom-cavity coupling $g$ on the memory efficiency. These parameters have a fundamental role in optimizing the efficiency, since the EIT transparency window is proportional to the rate $|\Omega_C|^2/g^2$ \cite{Villas_Boas10}. Thus, in order to store a Gaussian pulse with a given $FWHM$,  it is important to establish a relation between $g$ and $\Omega_C$ which ensures that the maximum width window of the EIT is fixed. From our simulations the best memory efficiency is obtained when $\Omega _{C}\simeq 2g/3$, for a pulse with $FWHM=1.0\mu\mathrm{s}$. This is due to the parameters that we are using in our simulations, which allows the pulse being inside the transparency window (in frequency domain). In contrast, if we consider a longer pulse (in time domain), it will be also inside the transparency window, however a large fraction of photons will be lost by transmission. 

\begin{figure}[h]
\centering
\includegraphics[width=1.0\linewidth]{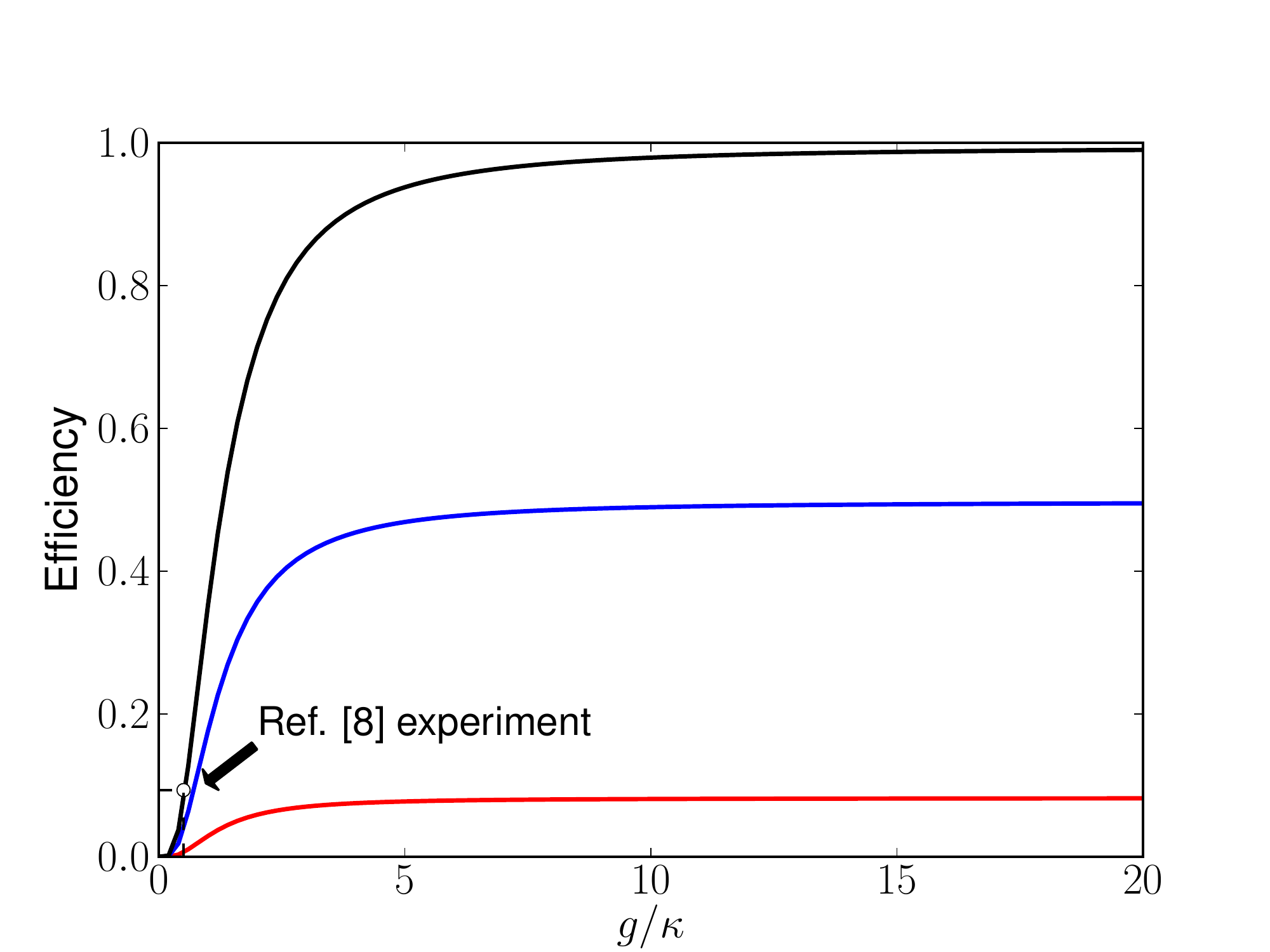}
\caption{Efficiency as function of  $g/\kappa$ for three different setups: asymmetric two-sided cavity (red solid line), symmetric two-sided cavity (blue solid line) and one-sided cavity (black solid line). We considered $\kappa=2.5\times 2\pi\mathrm{MHZ}$, being $\kappa_A=0.1\kappa_B$ and $\kappa_A=\kappa_B$ for asymmetric and symmetric two-sided cavities, respectively. In the case of the one-sided cavity, $\kappa=\kappa_A$ and $\kappa_B=0$. The parameters used here were: $\Gamma_{31}=\Gamma_{32}=0.6\kappa$, $\Omega_C = 2g/3$, $E_M = \sqrt{10^{-4}}\kappa$, $\zeta_1 = \zeta_2 = \zeta = 1.75$MHz, and $FWHM = 1.0\mu\mathrm{s}$. The dot marked on the black curve corresponds to the efficiency found in the single atom quantum memory experiment \cite{Rempe11}.}
\label{fig:3}
\end{figure}

From now on we will use a set of parameters according to that described in the previous paragraph for both cavity configurations, since we are interested in the highest efficiency value. In order to make a direct comparison between the three different setups in relation to the maximum efficiency value reached at each one, we plotted in the Fig. \ref{fig:3} the memory efficiency as function of the ratio $g/\kappa$, obtained for configuration $\mathrm{I}$, considering the asymmetric and symmetric cavity (red and blue solid curves, respectively) and configuration $\mathrm{II}$ (black solid curve). In our results we consider the total decay rate of the field equal for the three cases, $\kappa = 2\pi\times 2.5 \mathrm{MHZ}$. 

Firstly we investigated the memory efficiency considering the asymmetric two-sided cavity where the input pulse impinges on the left cavity mirror, which has a very high reflectivity compared to the right one. This experimental setup is exactly the one employed in \cite{Villas_Boas10} to observe cavity EIT with single atoms. In this setup, most of the probe laser is directly reflected while for the remainder photon fraction, one part is directly transmitted and another interacts with the atom. So, immediately we see that this scheme is not convenient for quantum memory purposes since only a small part of the light can enter the cavity and interact with the atom. We can note in Fig.\ref{fig:3} (red solid curve) that the efficiency saturates at about $8.5\%$, even for high values of $g$ coupling and, therefore, only a small fraction of the probe pulse can be storage in the atomic levels. This feature occurs due to the inevitable losses of the energy and information of the probe pulse due reflection (mainly) and transmission in this configuration. In this way,  although this setup has been very suitable for the observation of a narrow transmission window occurring at the two-photon resonance under cavity EIT conditions \cite{Villas_Boas10}, it is not convenient to perform a quantum memory.

Now we consider the configuration $\mathrm{I}$ but for cavity mirrors with equal transmission and reflection coefficients, i.e., considering a symmetric cavity (blue solid curve). Here, this experimental setup can be modelled assuming $\kappa_A=\kappa_B$, which means that light can leave the cavity from both sides. Before analysing the memory efficiency for this setup, we firstly present an analytical result to understand what could be expected here. To this end we analyse the steady state solution for the expressions of the external ($a_{out}(t)$, $b_{out}(t)$) and internal ($a(t)$) cavity modes described in the section \ref{sec:2} obtained via input-output formalism for a continuous driving regime, where $\phi_{in}$ is constant. The motion equation for the internal field operator $a(t)$ in our model is given by the Langevin equation \cite{Yamamoto86}:
\begin{equation}
\dot{a}(t)=-\kappa_T a(t)+\sqrt{2\kappa_A}\phi_{in}(t),
\label{eq:6}
\end{equation}
being $\kappa_T = \kappa_A+\kappa_B+\gamma$ the total damping rate of internal cavity field, where $\gamma$ represents the others field losses, such as the absorbed or scattered light by the atom. Thus, the equation (\ref{eq:6}) has the following stationary solution for the internal cavity mode:
\begin{equation}
\label{eq:7}
a^s(t)=\frac{\sqrt{2\kappa_A}}{\kappa_T}\phi_{in}.
\end{equation}      
Using the expression (\ref{eq:7}) and the input-output formalism that connects the external and internal cavity modes, the output cavity field in the stationary regime is given by:
\begin{subequations}
\begin{eqnarray}
\label{eq:8}
a^s_{out}&=&\frac{2\kappa_A-\kappa_T}{\kappa_T}\phi_{in}, \\
b^s_{out}&=&2\frac{\sqrt{\kappa_{A}\kappa_{B}}}{\kappa_T}\phi_{in}.
\end{eqnarray}
\end{subequations}

It is important to note that, for an empty symmetric cavity and a resonant probe field, corresponding to the case that $\gamma=0$ and $\kappa_A=\kappa_B=\kappa/2$, we obtain using the stationary solutions (\ref{eq:8}): $a^s_{out}=0$ and $b^s_{out}=\phi_{in}$. It means that there is no reflected light and the probe light is completely transmitted. So, one could expect a memory efficiency close to $100\%$ for this configuration as all the probe field cross the atom-cavity system, however this does not occur. Still in the stationary regime and when the input field can be absorbed by the atom, i.e., $\gamma \neq 0$, we can define a operator that describes the rate loss $\gamma$ as:
\begin{eqnarray}
\label{eq:9}
\sigma^s=\sqrt{2\gamma}a^s=\frac{\sqrt{2\eta}}{1+\eta}\phi_{in},
\end{eqnarray}
being $\eta=\gamma/2\kappa_A$.
This expression allows us to calculate the amount of the scattered or absorbed light by the atom, given by $|\langle\sigma^s\rangle|^2$, whose function is maximum at $\eta=1$. This implies that the maximum light absorption by the atom is $\frac{1}{2}|\phi_{in}|^2$, which results in a maximum expected efficiency for symmetric two-sided cavity of $50\%$. This analyses helps us to understand what can happen with our system, however it is valid for continuous driving. For coherent pulses interference processes could help to improve the efficiency but, as we show and explain below, this does not occur for this two-sided symmetric cavity and the efficiency is still limited to $50\%$.

Similarly to the asymmetric configuration, for the symmetric cavity there are losses due the transmission and reflection of the probe pulse. In this specific configuration, in which the rates $\kappa_A$ and $\kappa_B$ are equal, the transmission and reflection coefficients from the cavity have the same value, limiting the maximum efficiency value in $50\%$ as we see in Fig. \ref{fig:3} (blue solid curve). We can explain this maximum efficiency in the following way: in the regime $\Omega_C \ll g$, the transparency window of the EIT is narrower than the frequency width associated to the probe laser pulse, implying a high reflection of the pulse. In the opposite regime, $\Omega_C \gtrsim g$, the transmission of the system is very high and we are not able to turn the control field off fast enough to avoid the transmission of an expressive part of the input pulse, which is then immediately transmitted to the right side of the cavity before being able to be absorbed by the atom. We also tried to optimize the memory efficiency for this symmetric cavity using different shapes for the control field, even using the protocol by J. Dilley \textit{et al} \cite{Dilley12} for single photon pulses, but the we found that the maximum efficiency was always limited to $50\%$.

Ultimately remain in our analysis the maximum efficiency value for one-sided cavity case (black solid curve). This setup in our theoretical model implies in the condition $\kappa_B = 0$. In this configuration we assume that the right mirror reflects $100\%$ of the light and then, the photons fraction retrieved leaves the cavity from the same side which the input pulse is sent. From the steady state analyses performed above for the continuous driving, but now considering $\kappa_B = 0$, we see from the stationary solutions (\ref{eq:8}) and (\ref{eq:9}): $a^s_{out} \simeq - \phi_{in}$ and $\sigma^s \simeq 0$ in the limit of $\gamma \gg \kappa_A$. This means that this simple analyses is not enough to explain how the memory efficiency can reach values close to $100\%$ in the one-sided cavity setup, thus requiring interference process to understand it.

We observe in Fig.\ref{fig:3} that for the same set of parameters that optimizes the efficiency for the others configurations, considering the one-sided cavity, the efficiency has its maximum value increased considerably compared to them. For this specific setup the maximum efficiency value is close to $100\%$, for sufficiently high values of the coupling constant $g$. The white dot evidences the efficiency value experimentally measured in \cite{Rempe11} ($9.3\pm 1$)$\%$ when the $g\approx 1.09\kappa$. The high efficiency value reached in the strong coupling regime for the one-sided cavity occurs due to the interference process between the field immediately reflected when the input pulse impinges on the cavity mirror, and the field that enters the cavity and then is transmitted to the outside, by the same side, after one round trip. In this way, from the experimental point of view both fields, reflected and transmitted, go through the same path becoming indistinguishable to a detector leading to an interference process, significantly increasing the memory efficiency. 

In relation to the one-sided cavity configuration, is important to mention that in the past years one has theoretically demonstrated that for this specific setup and single photon pulses the value of quantum memory efficiency can reach near $100\%$, if the impedance matching condition is fulfilled \cite{Lukin00, Dilley12}. This condition requires an specific time dependency for the control field $\Omega_C(t)$, derived from input-output theory, such that the reflected and transmitted fields from the cavity interfere destructively, completely annihilating each other. Using this approach, J. Dilley \emph{et al.} \cite{Dilley12}, showed that for a sufficiently high cooperativity $C = g^2/\kappa\Gamma$, being $\Gamma$ the total spontaneous-emission rate of the excited state of the atom, a memory efficiency arbitrarily close to $100\%$ is obtained. It is relevant to mention that in our simulation we were not able to apply phase-matching conditions proposed by J. Dilley \emph{et al.} \cite{Dilley12} since we are using weak coherent pulses. However, even using a simple form of the control field (\ref{eq:3}), we were able to achieve an efficiency close to $100\%$ for the quantum memory, simply by choosing the best parameters for the system and adiabatically turning off and on the control field. For smooth pulses and symmetric in time (such as in our case that we consider a Gaussian pulse) some parameters such as the relation between the time that the input pulse arrives and time that the control field is adiabatically turned off, are more relevant in order to obtain high efficiency values than the actual shape of the control field derived from the impedance matching condition \cite{Lukin00}. In fact, the maximum value of the efficiency calculated using this specific form for $\Omega_C$ is extremely sensitive to any minimal change in the field shape, thus making an experimental challenge the incidence of a pulse with a shape very specific and precise. 


\textit{Optical Transistor:} We are also interested in investigating the performance of an optical transistor based on EIT phenomenon. So, here we analyse and compare the transmission spectrum of the system for both experimental setups under EIT conditions. As mentioned previously the asymmetric two-sided cavity is exactly the experimental setup employed in \cite{Villas_Boas10} to observe cavity EIT with single atoms. In that experiment a single atom was able to control the passage of light through the atom-cavity system, allowing this setup being named as "single-atom transistor for light" \cite{Parkins10}. In this scheme, most of the probe laser is directly reflected implying that it can not work out perfectly as an optical transistor since even in the absence of the atom most of light is reflected.

\begin{figure}[hbt]
\centering
\includegraphics[width=1.0\linewidth]{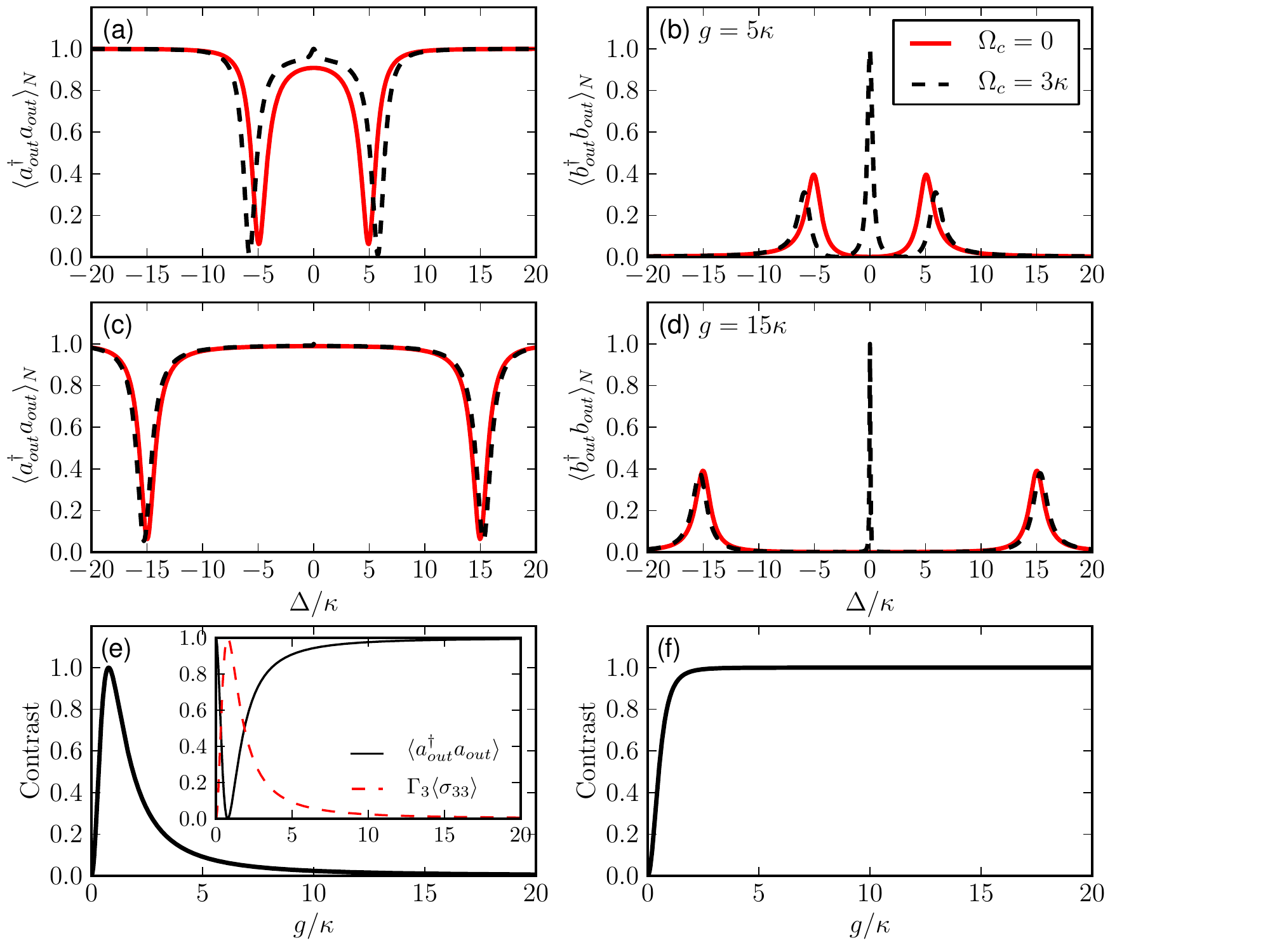}
\caption{Normalized cavity transmission for cavity EIT as function of $\Delta/\kappa$, considering (a) and (b) $g = 5\kappa$ and (c) and (d) $g = 15\kappa$. In the red solid curve $\Omega_C = 0$, and in the black dashed curve $\Omega_C = 3\kappa$. (e) and (f) show the contrast given by the difference between these two curves at resonance, $\Delta = 0$. The inset shows the mean number of photons from cavity (black solid curve) and the scattered light by the atom (dashed red curve) as function of the ratio $g/\kappa$, considering the configuration $\mathrm{II}$. The left panels are associated to an one-sided cavity system, while the right panels correspond to two-sided cavity system.}
\label{fig:6}
\end{figure}

To have an ideal optical transitor, which can control the passage of light through the application of a second light field, we must look for a setup which allows us to have a perfect contrast between the transmission when the control field is on and off. This contrast can be achieved using the configuration $\mathrm{I}$ with symmetric cavity, since it allows for a $100\%$ transmission of light in the absence of atom or when there is an atom inside it in the EIT regime, i.e., for $\Omega_{C}$ non null. Such feature can be derived from the stationary solutions (\ref{eq:8}). On the other hand, when we have an atom inside the cavity in the strong atom-field coupling regime and the control field is off, the transmission of a probe field goes to zero due to the normal mode splitting of the atom-cavity system. Thus, placing a three-level atom inside a symmetric cavity the transmission can be either maximum ($100\%$) for $\Omega_{C} \neq 0 $ or null, i.e., the probe field is completely reflected, for $\Omega_{C} = 0$. Thus, here we have a perfect (single-atom) optical transistor. The transmission measurements considering the control field on or off, evidences the feasibility of the implementation of an optical transistor using two-side symmetric cavity setup. On the other hand, the one-sided cavity is not suitable to observe cavity EIT in the transmission spectrum, as we explain below. 

In Fig. \ref{fig:6}(a)-(d) the theoretical curves of the transmission spectrum normalized by the input field at resonance are plotted for different atom-cavity coupling regimes as function of the normalized detuning $\Delta/\kappa$, considering $\Omega_C=0$ (red solid line) and $\Omega_C=3\kappa$ (black dashed line). For a continuous driving and $\Omega_C=0$ the atomic population goes asymptotically to state $|2\rangle$ and then we would end up with an empty cavity situation. To avoid this and to simulate a two-level atomic system, for  $\Omega_C=0$ we must also artificially adjust $\Gamma_{32} = 0$. In the panels (a) and (b), $g = 5\kappa$ and, in the panels (c) and (d), $g = 15\kappa$. In Figs. \ref{fig:6}(e) and (f) is plotted the contrast of the normalized mean number of photons outside the cavity as a function of the constant coupling $g$ at resonance, $\Delta=0$. This amount is given by the difference between the mean number of photons outside the cavity when the control field is turned on ($\Omega_C = 3\kappa$) and when it is absent ($\Omega_C = 3\kappa$). In Fig. \ref{fig:6} the left panels (a, c and e) correspond to one-sided cavity system (field operator $a_{out}$) and, the right panels (b, d and f) are associated to the system constituted by two-side symmetric cavity (field operator $b_{out}$). The important feature that we can observe in these results is that the transmission peak at resonance, which in turn is a signature of transparency window, does not appear in the transmission spectrum of the configuration $\mathrm{II}$. Therefore, the EIT phenomenon can not be properly observed in experiments based on transmission measurements in the adequate experimental setup for optical quantum memory (setup $\mathrm{II}$). The big difference between the behaviour of these two configurations, in the EIT regime, is entirely clear in the contrast measure. In Fig. \ref{fig:6}(f) the contrast goes to unity as the coupling $g$ increases, evidencing that the system is completely transparent to the probe laser when $\Omega_C$ is turned on. Otherwise, for one-sided cavity the contrast exhibits a peak at $g/\kappa \approx0.8 \kappa$, whose maximum is $1.0$ and for high values of $g$ it tends to zero. This occurs since in the strong coupling regime the light is reflected if $\Omega_C=0$ or transmitted if $\Omega_C\neq0$. In this way, for cavity EIT experiments performed in one-sided cavity it is not possible to distinguish the reflected and transmitted fields using intensity measurements and thus, as it can be seen in the Fig. \ref{fig:6}(c), both curves overlap completely, resulting in null contrast. Then, it becomes more complicated to this setup be used to implement an optical transistor. 

One way to distinguish the reflect and transmitted fields from the one-sided cavity system is through the phase difference between them, which is induced by the control laser. If the control laser is turned off, only the atomic levels $|1\rangle$ and $|3\rangle$ participate of the dynamics and, according to the Jaynes-Cummings model, the photons that impinge onto the system at resonance $\Delta = 0$, do not enter the cavity. This happens because, due to the mode splitting caused by the atom-cavity coupling,  the probe laser that was resonant with the transition $|1\rangle-|3\rangle$ for the empty cavity is now directly reflected by the left mirror and its phase $\phi_r$ is shifted of $\pi$. Conversely if the control laser couples resonantly the $|2\rangle-|3\rangle$ transition, the probe laser can enters the cavity interacting with the atom and then being transmitted with $\phi_t$ phase, without experiencing any change in its phase. Therefore, in this experimental setup the control laser has an important role to induce a phase difference between the reflected and transmitted fields, such that $|\phi_t - \phi_r| = \pi$, when the probe laser is resonant with the cavity mode. In fact, this system can be used to produce a phase gate with a classical field (control field) inducing a $\pi$ phase in a quantum field, for example a superposition of zero and one photon, or superposition of different polarization states of the field: when the control field is off the quantum field acquires a $\pi$ phase shift. On the other hand the quantum field does not acquire any phase shift when the control field is on \cite{halyne2015}.

For $g=0.8\kappa$, all the light from the probe field goes into the cavity and it is completely absorbed by the atom, which scatters the light. In this situation there is no reflected light and then we have an interesting effect where the atom is able to scatter $100\%$ of the light incident on the atom-cavity system. According to our simulations, this effect depends not only on the atom-field coupling $g$ but also on the probe field intensity: for stronger probe field the complete scattering happens for stronger $g$. This specific value of the atom-field coupling $g$ where all the light is completely scattered can also be obtained using the stationary solutions given by equations (\ref{eq:8}) and (\ref{eq:9}), if we consider $\gamma=\kappa_A$. Considering this specific condition for one-sided cavity where $\kappa_A=\kappa$, we found $a_{out}=0$, i.e., the outside field is zero and $\sigma^s=\phi_{in}$. Due to the presence of the atom inside the cavity we can derive an effective field decay which is given by $\gamma=g^2/\Gamma_{31}$ \cite{Werlang08, Prado11}. Thus, as the light is completely scattered by the atom when the condition $\gamma=\kappa$ is fulfilled, we obtained $g=\sqrt{\Gamma_{31}\kappa}$. In our numeric simulations we assume $\Gamma_{31}=0.6\kappa$, which implies $g\sim 0.8\kappa$. Therefore, this result obtained from the stationary expressions is in complete agreement with the numerical results showed in the inset of the Fig.\ref{fig:6}(e).

\section{Conclusions}
\label{sec:4}

Here we have presented theoretical results concerning the implementation of quantum memory and optical transistor in cavity QED with a single trapped atom based on electromagnetically induced transparency. This atom-cavity system was already used for the generation of single photons, to implement some quantum logic gates, quantum  memory and others applications \cite{Reiserer15}, which turns this system very attractive for quantum information processing. Depending on the mirrors configuration we can define two different experimental setups: configurations $\mathrm{I}$ two-sided (asymmetric or symmetric) or $\mathrm{II}$ one-sided cavity. We have shown that the asymmetric cavity employed in \cite{Villas_Boas10} to observe single atom cavity EIT is neither convenient to perform quantum memory or optical transistor. Still in the two-sided cavity setup, the symmetrical one allows for a perfect contrast of the transmitted field between the situation when the control field is on and off, for strong atom-field coupling $g$, thus being suitable for the implementation of an ideal optical transistor. However, this setup does not allow for high quantum memory efficiencies, being limited to $50\%$ for strong $g$. On the other hand, the configuration $\mathrm{II}$ is very suitable for quantum memory applications, allowing efficiencies close to $100\%$, but it can not be used as an optical transistor since both fields, reflected (control field off) and transmitted (control field on) leave the cavity by the same side, being indistinguishable in transmission measurements. In order to observe the EIT phenomenon in one-sided cavity system with coherent probe fields it is necessary another kind of experiment, for example carrying out phase measurements which would need additional optical linear devices and local oscillators, for instance, which would introduce other error sources. 

\section{\label{ackn}Acknowledgements}
\label{sec:5}

R.R.O., C. J. V.-B. and H. S. Borges acknowledges support from CNPq and FAPESP (Proc.
2012/00176-9, 2013/04162-5 and 2014/12740-1), and the Brazilian National Institute for
Science and Technology of Quantum Information (INCT-IQ). We also thank the fruitful discussions with Stephan Ritter.

\end{document}